\documentclass[11pt]{article}
\usepackage{graphicx}
\usepackage{epsf}

\newcommand{\BaBarPubYear}       {08}

\newcommand{\BaBarConfNumber}    {001}
\newcommand{\SLACPubNumber} {13325}
\newcommand{\LANLNumber} {0807.4187}

\def\Journal#1#2#3#4{{#1} {\bf #2}, #3 (#4)}

\def\NIMA{{\em Nucl. Instrum. Methods} A}

\input babarsym

\def\bmunu     {\ensuremath{\Bp \to \mup \num}\xspace}
\def\btaunu     {\ensuremath{\Bp \to \taup \nut}\xspace}
\def\blnu     {\ensuremath{\Bp \to \ellp \nul}\xspace}
\def\benu       {\ensuremath{B^{\pm} \to \ep \nue}\xspace}

\long\def\inst#1{\par\nobreak\kern 4pt\nobreak
    {\it #1}\par\vskip 10pt plus 3pt minus 3pt}
\begin{document}

{\pagestyle{empty}

\begin{flushright}
\babar-CONF-\BaBarPubYear/\BaBarConfNumber \\
SLAC-PUB-\SLACPubNumber \\
arXiv:\LANLNumber \\
July, 2008 \\
\end{flushright}

\par\vskip 5cm

\begin{center}
\Large \bf Search for $B^+\rightarrow\mu^+\nu_{\mu}$ with inclusive reconstruction at \babar
\end{center}
\bigskip

\begin{center}
\large The \babar\ Collaboration\\
\mbox{ }\\
\today
\end{center}
\bigskip \bigskip

\begin{center}
\large \bf Abstract
\end{center}
We search for the purely leptonic decay $B^{\pm} \rightarrow \mu^{\pm} \nu_{\mu}$  in the full 
\babar\  dataset, having an integrated luminosity of approximately 426 fb$^{-1}$. 
We adopt a fully inclusive approach, where the signal candidate is identified by the highest momentum lepton 
in the event and the companion $B$ is inclusively reconstructed without trying to identify its decay products.  
We set a preliminary upper limit on the branching fraction of
$\BR(B^{\pm} \rightarrow \mu^{\pm} \nu_{\mu}) < 1.3 \times 10^{-6}$ at the 90\% confidence level, using a Bayesian approach. 

\vfill
\begin{center}
Presented at the 
International Conference On High Energy Physics (ICHEP 2008),
7/29---8/5/2008, Philadelphia, USA
\end{center}

\vspace{1.0cm}
\begin{center}
{\em Stanford Linear Accelerator Center, Stanford University, 
Stanford, CA 94309} \\ \vspace{0.1cm}\hrule\vspace{0.1cm}
Work supported in part by Department of Energy contract DE-AC03-76SF00515.
\end{center}

\newpage
}

\begin{center}
\small

The \babar\ Collaboration,
\bigskip

%
B.~Aubert,
M.~Bona,
Y.~Karyotakis,
J.~P.~Lees,
V.~Poireau,
E.~Prencipe,
X.~Prudent,
V.~Tisserand
\inst{Laboratoire de Physique des Particules, IN2P3/CNRS et Universit\'e de Savoie, F-74941 Annecy-Le-Vieux, France }
J.~Garra~Tico,
E.~Grauges
\inst{Universitat de Barcelona, Facultat de Fisica, Departament ECM, E-08028 Barcelona, Spain }
% ...

\end{center}\newpage

\setcounter{footnote}{0}

\section{INTRODUCTION}
\label{sec:Introduction}

\begin{figure}[htb]
\begin{center}
\includegraphics*[width=0.40\textwidth]{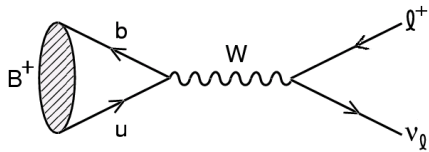}
\label{fig:diagr} 
\caption{SM annihilation diagram for $B^{\pm} \rightarrow l^{\pm} \nu_{\mu}$.}
\end{center}
\end{figure}

In the Standard Model (SM), the purely leptonic $B$ decays \blnu 
( $\ell = e, \mu, \tau$ ) (charge conjugation is implied troughout the paper)
proceed through the annihilation of the two quarks in the 
meson to form a virtual $W$ boson (Fig.~\ref{fig:diagr}). The branching ratio can be cleanly 
calculated in the SM,    

\begin{equation}
 \BR(B^+ \rightarrow \ell^+ \nu_{\ell}) = \frac{G_{F}^{2} m_{B} m_{\ell}^{2}} {8\pi} 
 \biggl( 1- \frac{m_{\ell}^{2}}{m_{\B}^{2}} \biggr)^{2} f_{B}^{2} |V_{ub}|^{2} 
 \tau_{\B},
\end{equation}

where $G_F$ is the Fermi coupling constant, $m_{\ell}$ and $m_B$ are 
the lepton 
and $B$ meson masses, and $\tau_B$ is the $B^+$ lifetime. The decay rate 
is sensitive to the Cabibbo Kobayashi Maskawa 
matrix element $V_{ub}$ and the $B$ decay constant $f_{\B}$ which describes 
the overlap of the quark wave functions within the meson. Currently, 
the uncertainty on $f_B$ is one of the main
factors limiting the determination of $V_{td}$ from precision $B^0\bar{B}^0$ 
mixing measurements. Given a measurement of 
$V_{ub}$ from semileptonic decays such as $B\rightarrow\pi\ell\nu$, 
$f_B$ could be extracted from  a measurement of the 
\blnu branching ratio. 

The SM estimate of the branching ratio for 
\btaunu is $(1.59\pm0.40)\times 10^{-4}$ assuming 
$\tau_B$ = 1.638$\pm$0.011 ps, $V_{ub}$ = (4.39$\pm$0.33)$\times 10^{-3}$\cite{Barberio:2006bi} 
determined from inclusive charmless semileptonic $B$ decays and $f_B$ = 216$\pm$22 MeV~\cite{Gray:2005ad}
from lattice QCD calculation. 
Due to helicity suppression, \bmunu and \benu are 
suppressed by factors of 225 and $10^{7}$ respectively, leading to   
branching ratios of $\BR(B^{\pm} \rightarrow \mu^{\pm} \nu_{\mu} ) \simeq 4.7 \times 10^{-7}$ and
$\BR(B^{\pm} \rightarrow e^{\pm} \nu_{e}) \simeq 1.1 \times 10^{-11}$. 

Purely leptonic $B$ decays are sensitive to physics beyond the SM due to possible
insertion of New Physics (NP) heavy states in the annihilation process.
Charged Higgs boson effects may greatly enhance or suppress the 
branching ratio in certain two Higgs doublet models \cite{Hou:1992sy}. Similarly, this decay
may be enhanced through mediation by leptoquarks in the Pati-Salam model of quark-lepton
unification \cite{Valencia:1994cj}. 

Moreover, as in annihilation processes the longitudinal component of the vector boson is directly 
involved, this decay allows a  direct test of Yukawa interactions in and beyond the SM. In particular, 
in a SUSY scenario at large $\tan \beta$ ($O(m_t/m_b) >> 1$), non-standard effects in helicity-suppressed 
charged current interactions are potentially observable, being strongly $\tan \beta$ dependent:
\begin{eqnarray}
\BR(B^{\pm} \rightarrow l^{\pm} \nu_{l}) \approx \BR(B^{\pm} \rightarrow l^{\pm} \nu_{l})_{\rm{SM}} \times \Big( 1-  \tan \beta^2 m_B^2/M_H^2 \Big)^2 .
\end{eqnarray}
\newline
Recently, Belle had a first evidence of a purely leptonic \B decay.
With 414 \invfb, Belle finds~\cite{Ikado:2006un}
\begin{equation}
 \BR(B^+\rightarrow\tau^+\nu_{\tau}) = \Big(1.79^{+0.56}_{-0.49}(\rm stat)^{+0.46}_{-0.51}(\rm syst)\Big) \times 10^{-4},
\end{equation}
at 3.5 $\sigma$ significance. The most recent \babar\ result on this channel uses an integrated luminosity of 346 fb$^{-1}$, corresponding to 383 million of $B \bar{B}$ pairs,
and sets an upper limit (UL) at $90\%$ of confidence level on the branching ratio of $ \BR(B^+\rightarrow\tau^+\nu_{\tau}) < 1.7\times 10^{-4}$\cite{Aubert:2007}
and a central value  $ \BR(B^+\rightarrow\tau^+\nu_{\tau}) = (1.2 \pm 0.4 \stat \pm 0.3 ({\rm bkg}) \pm 0.2 \syst)\times 10^{-4}$\cite{Aubert:2007xj}.

\babar\ has published a result on $ \BR(B^+\rightarrow\mu^+\nu_{\mu} )$  with 81 fb$^{-1}$ and set an UL at 90 $\%$  confidence level
of $ \BR(B^+\rightarrow\mu^+\nu_{\mu} ) < 6.6 \times 10^{-6}$~\cite{Aubert:2004yu}.
The current best published upper limits at 90 $\%$  confidence level on \bmunu and \benu are from Belle Collaboration on 235 \invfb~\cite{Satoyama:2006xn}
\begin{eqnarray}
  \BR(B^+\rightarrow\mu^+\nu_{\mu}) &<& 1.7 \times 10^{-6} , \nonumber\\
  \BR(B^+\rightarrow e^+\nu_{e}) &<& 9.8 \times 10^{-7}.
\end{eqnarray}

\section{THE \babar\ DETECTOR AND DATASET}
\label{sec:babar}

This analysis is based on the data collected with the \babar\ detector~\cite{NIM}
at the PEP-II storage ring. The sample corresponds to an integrated luminosity
of 426\invfb at the $\Upsilon(4S)$ resonance, consisting of about 447
millions of \BB pairs, and  44\invfb accumulated 
at a center-of-mass (CM) energy about 40 \mev below the \FourS resonance.
Off-resonance data are used as cross-checks for continuum $q \bar{q}$ (\q = \u, \d, \s, and \c) and $\tau^+ \tau^-$
on-resonance events.
In particular, given the variation of muon identification in time due to
detector differences and changes, we considered the total dataset divided into
data-taking periods (runs).

Charged track reconstruction is provided by a Silicon Vertex Tracker (SVT)
and a Drift Chamber (DCH) operating in a 1.5-T magnetic field. Particle
identification is based on the energy loss \dedx in the tracking system and the
Cherenkov angle in an internally reflecting ring-imaging Cherenkov detector.
Photon detection is provided by a CsI(Tl) Electromagnetic Calorimeter (EMC). Muons
and neutral hadrons are identified by Resistive Plate Chambers and Limited Streamer Tubes
in the Instrumented Flux Return (IFR) detector.

A GEANT4-based~\cite{geant} Monte Carlo (MC) simulation is used to model the
detector response and test the analysis technique.  A sample of about 28 million  
simulated \BpBm events where $B^+$ decays to $\mu^+ \nu_{\mu}$ and the \Bm decays generically is studied
to evaluate the efficiency for the signal. 
Background sources considered include \epem\to\BB, \epem\to\qqbar 
(\q = \u, \d, \s, and \c), and \epem\to\tautau in quantities comparable 
to  three times  (\BB), twice  ($c \bar{c}$)
and once ($uds$,\tautau) the actual dataset luminosity.

\section{ANALYSIS METHOD}
\label{sec:Analysis}

\begin{figure}[htb!]
\begin{center}
\includegraphics*[width=0.45\textwidth]{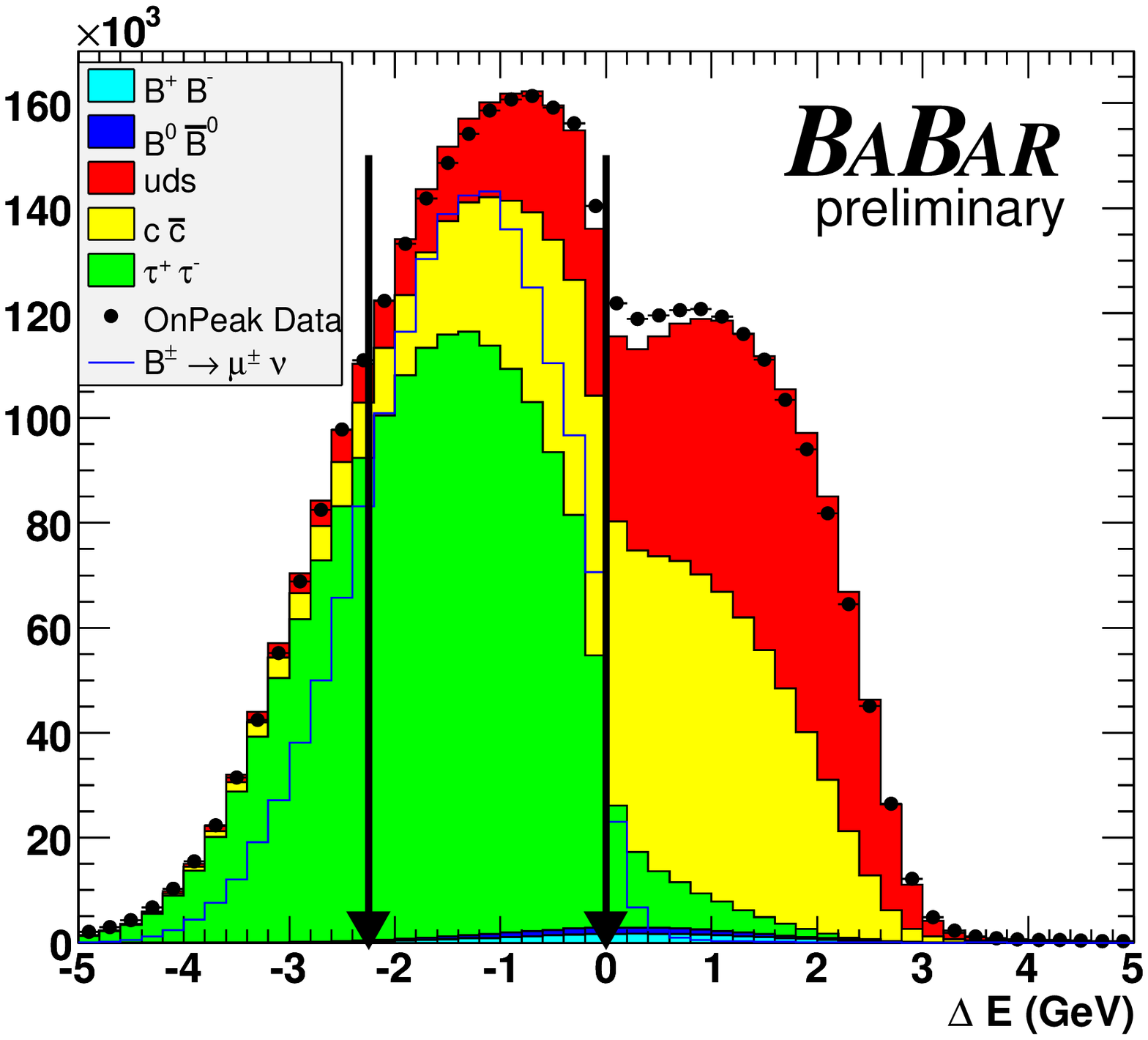}
\includegraphics*[width=0.45\textwidth]{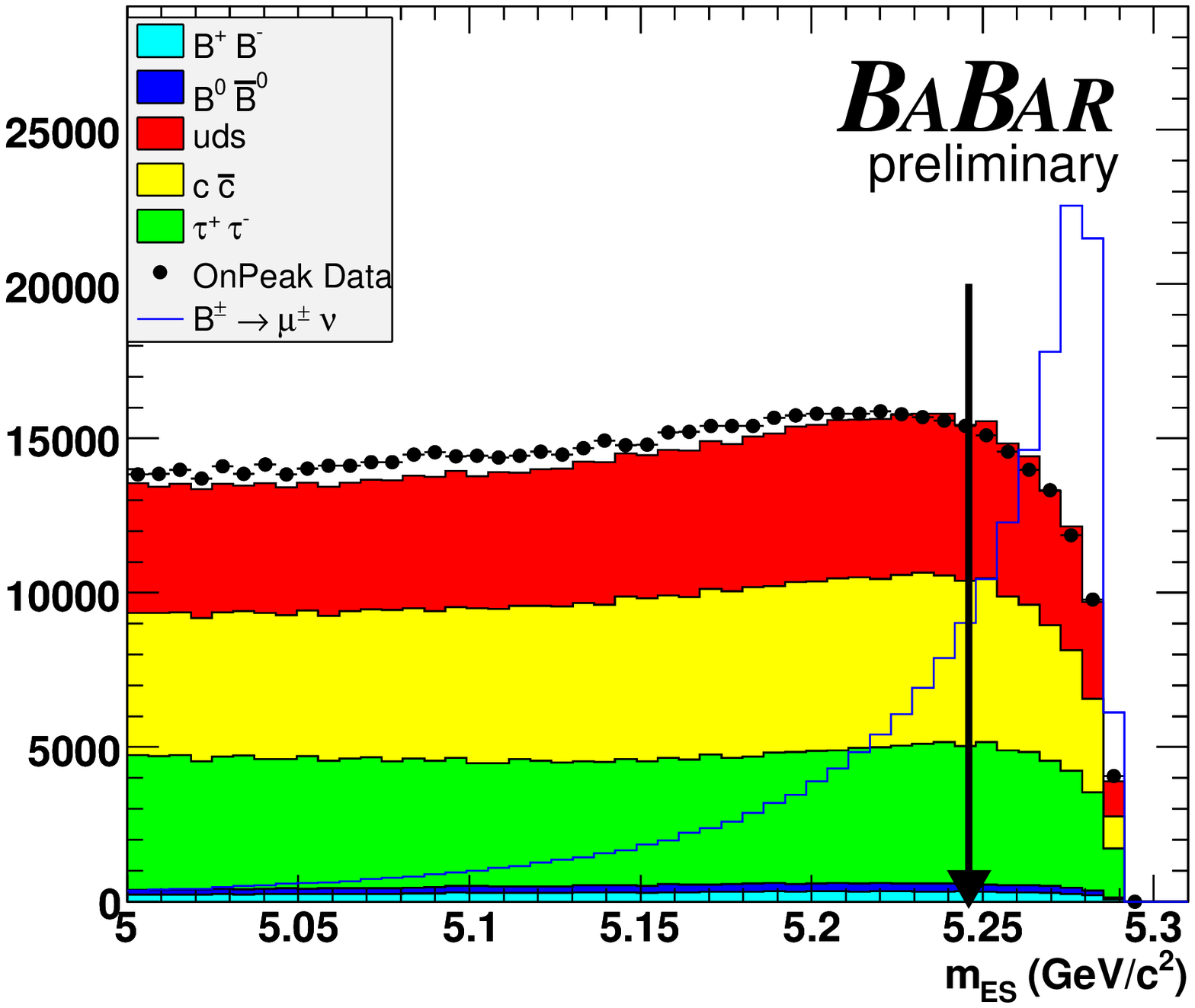}
\caption{ Distribution of $\Delta E$ and $m_{ES}$ after the muon selection:
  signal in blue histogram, data in black dots and background events are stacked on top of each other:  
  $uds$ in red, $c\bar{c}$ in yellow, $\tau^+\tau^-$ in green, $B^0 \bar{B}^0$ in dark blue and 
  $B^+ B^-$ in light blue. The arrows indicate the requirements set on these variables.}
\label{fig:de_mes} 
\end{center}
\end{figure}

 \bmunu is a two-body decay so the 
muon must be mono-energetic in the \B rest frame. The momentum $p^*$ of
the muon in the $B$ rest frame is given by
\begin{equation}
 p^* = \frac{m^2_B - m^2_{\mu}}{2m_B} \approx \frac{m_B}{2} \approx 2.46 {\rm  GeV}.
\end{equation}
where $m_B$ is the $B$ mass and $m_{\mu}$ is the muon mass.
At \babar\, the CM frame is a good approximation to the \B 
rest frame, so we initially select well-identified muon candidates with
momentum $p_{\rm{CM}}$ between 2.4 and 3.2 \gevc in the CM frame. Since the neutrino produced in the 
signal decay is not detected, any other charged tracks or neutral deposits in a signal 
event must have been produced by the decay of the companion (tag) \B. Therefore, the tag 
\B can be reconstructed from the remaining visible energy in the event.
Signal decays can then be selected using the kinematic variables \DeltaE and energy-substituted mass, \mes, 
defined by
\begin{equation}
 \DeltaE = E_{B}-E_{\rm beam},
\end{equation}
and
\begin{equation}
 \mes = \sqrt{E_{\rm beam}^{2}-|\vec{p}_B|^{\;2}},
\end{equation}
where $\vec{p}_B$ and $E_B$ are the momentum and energy of the reconstructed 
tag \B candidate in the CM frame and $E_{\rm beam}$ is the beam energy in the CM frame.
We include all neutral calorimeter clusters with 
cluster energy greater than 30 \mev. Particle identification is applied to the charged tracks 
to identify electrons, muons, kaons and protons in order to apply the most likely mass 
hypothesis to each track and thus improve the \DeltaE and \mes resolution. 
Events with additional identified leptons are discarded to discriminate against events containing additional 
neutrinos. 
For signal events in
which all decay products of the other \B are reconstructed, we expect the $\Delta E$ distribution 
to peak near zero and \mes to peak near $m_B$. In reality, due to the inclusive
nature of our analysis, we often fail 
to reconstruct all the decay products so that the $\Delta E$ distribution develops a negative tail 
while the $m_{ES}$ distribution exhibits a tail below the \B mass.  For $uds$ and 
$c\bar{c}$ backgrounds, $\Delta E$ is shifted significantly greater than zero 
since we attribute too much energy to the opposite hemisphere decay. 
$\Delta E$ is negative for $\tau^+\tau^-$ decays due to missing 
neutrinos. 
Figure~\ref{fig:de_mes} shows the distributions of \DeltaE and \mes for the 
on-peak data, background MC and signal MC after muon candidate selection. 

In order to obtain data/MC agreement, we extract the $uds$, $c \bar{c}$ and $\tau^+ \tau^-$ MC
normalization coefficients from a fit to the $\Delta E$ data distribution, keeping the $b \bar{b}$ component fixed.
The requirements on tag $B$ kinematical variables is optimized with the figure of merit $\epsilon_{\rm{sig}}/\sqrt{N_{\rm{bkg}}}$ where
$\epsilon_{\rm{sig}}$ is the signal efficiency and $N_{\rm{bkg}}$ the number of background events.
We require the tag $B$ $\Delta E$ and $m_{ES}$ to be within -2.25 $< \Delta E <$ 0 GeV and $m_{ES}>5.246$ GeV/c$^2$.

Once the tag \B is reconstructed, we refine the estimate of the muon momentum in the \B rest
frame ($p^*$). We use the momentum direction of the tag \B and assume a total momentum of 320 \mevc in the CM frame (from the decay 
of the \FourS\to\BB) to boost the muon candidate into the reconstructed \B rest frame. 

Backgrounds may arise from any process producing charged tracks in the 
momentum range of the signal, particularly if the charged tracks are  
muons. The two most significant backgrounds are \B semileptonic decays involving
$b\rightarrow u\mu\nu_{\mu}$ transitions where the endpoint of the muon 
spectrum approaches that of the signal, and non-resonant \qqbar (continuum) events 
where a charged pion 
is mistakenly identified as a muon. In the continuum events, there must also be 
significant missing energy due to detector acceptance, neutral hadrons, or additional neutrinos 
that mimic the signature of the expected neutrino.  

Continuum backgrounds are suppressed using event shape variables. The light-quark events
tend to produce a jet-like event topology as opposed to \BB events which tend to be 
more isotropically distributed in space. Several topological variables have been considered and five have been found 
to be the most discriminating, using an appropriate cocktail of different data-taking periods.
These variables are combined in a Fisher discriminant~\cite{fisher}:
the normalized second Fox-Wolfram moment $R_2$~\cite{Fox:1978vu}, calculated using
all charged tracks and neutral clusters in the event; the ratio of the second 
and zeroth Legendre Polynomials $L_2/L_0$, where all tag \B daughters momenta in the CM frame
are included and the angle is measured with respect to the lepton candidate momentum;
the cosine of the angle of the expected 
signal neutrino in the lab frame (as determined from the lepton candidate); the lepton transverse
momentum in $\Upsilon(4S)$ frame; the sphericity of the event. The Fisher coefficients are 
optimized run-by-run.
A cut is applied on the
Fisher discriminant and is thus optimized for each run separately in order to have better performance. The
efficiency of this cut is in the range 16$\%$ to 32$\%$ for signal events , 5$\%$ to 16$\%$ for $b \bar{b}$
events and less than 0.5$\%$ for continuum events.

\begin{figure}[h]
\begin{center}
\includegraphics*[width=0.40\textwidth]{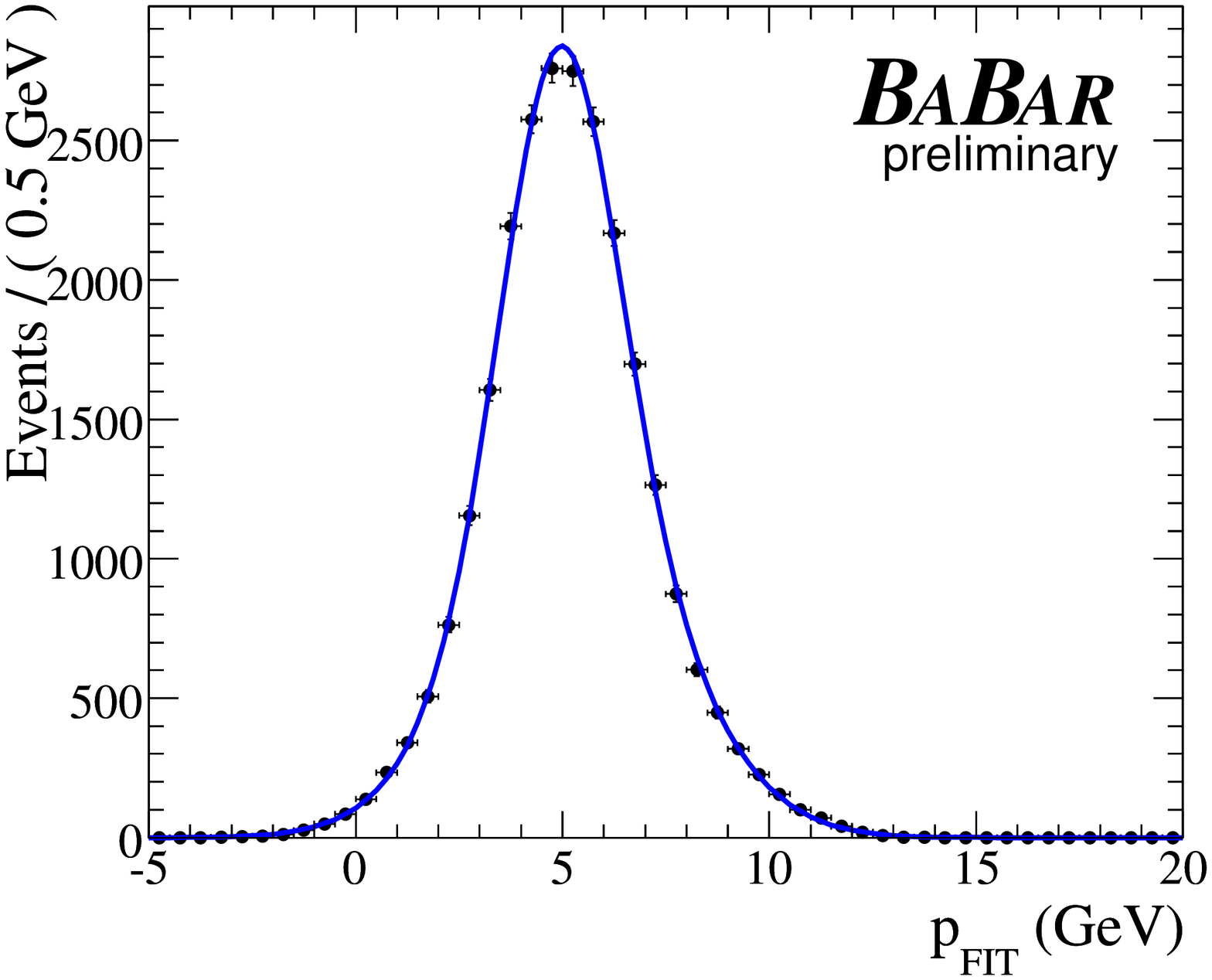}
\includegraphics*[width=0.40\textwidth]{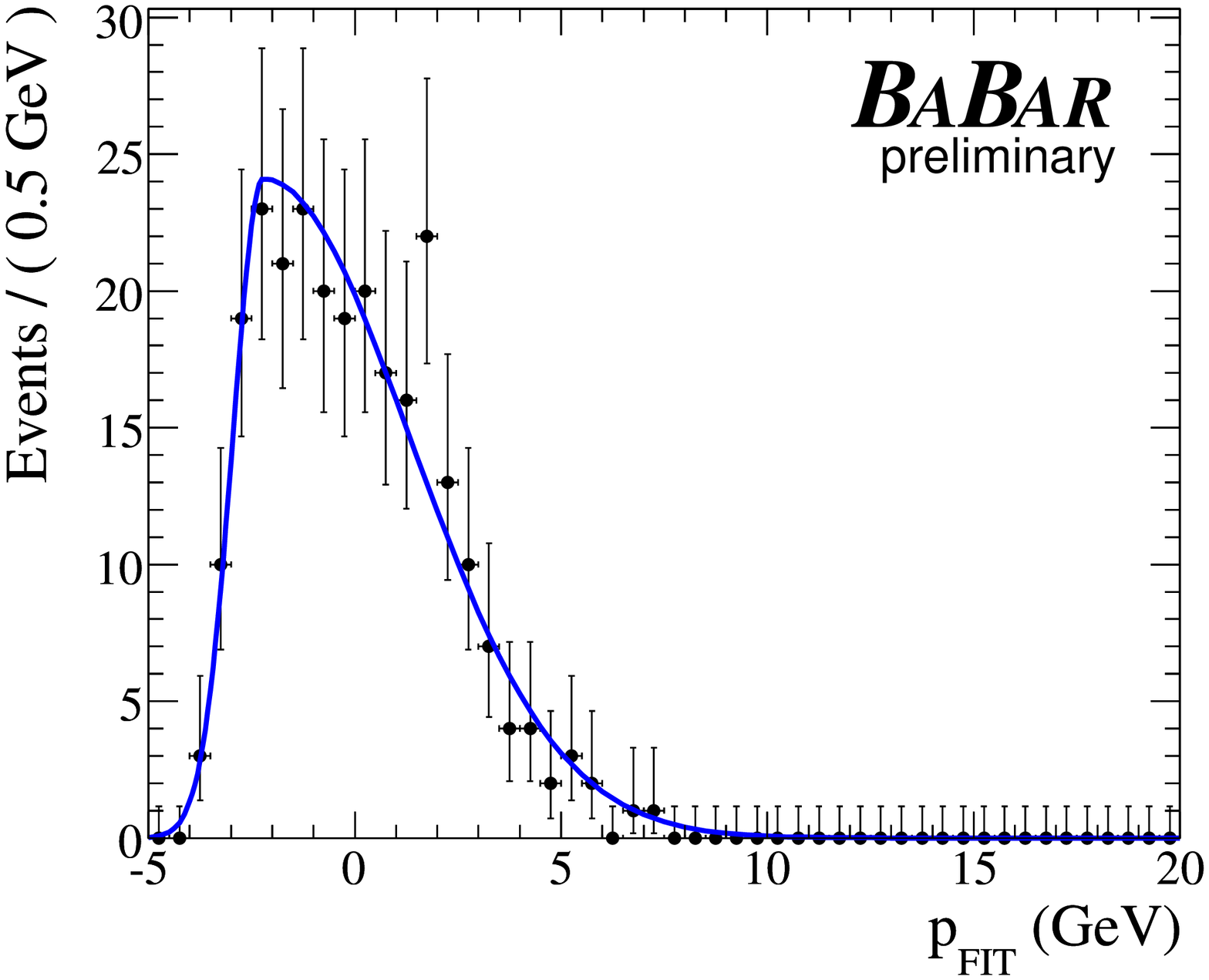}
\includegraphics*[width=0.40\textwidth]{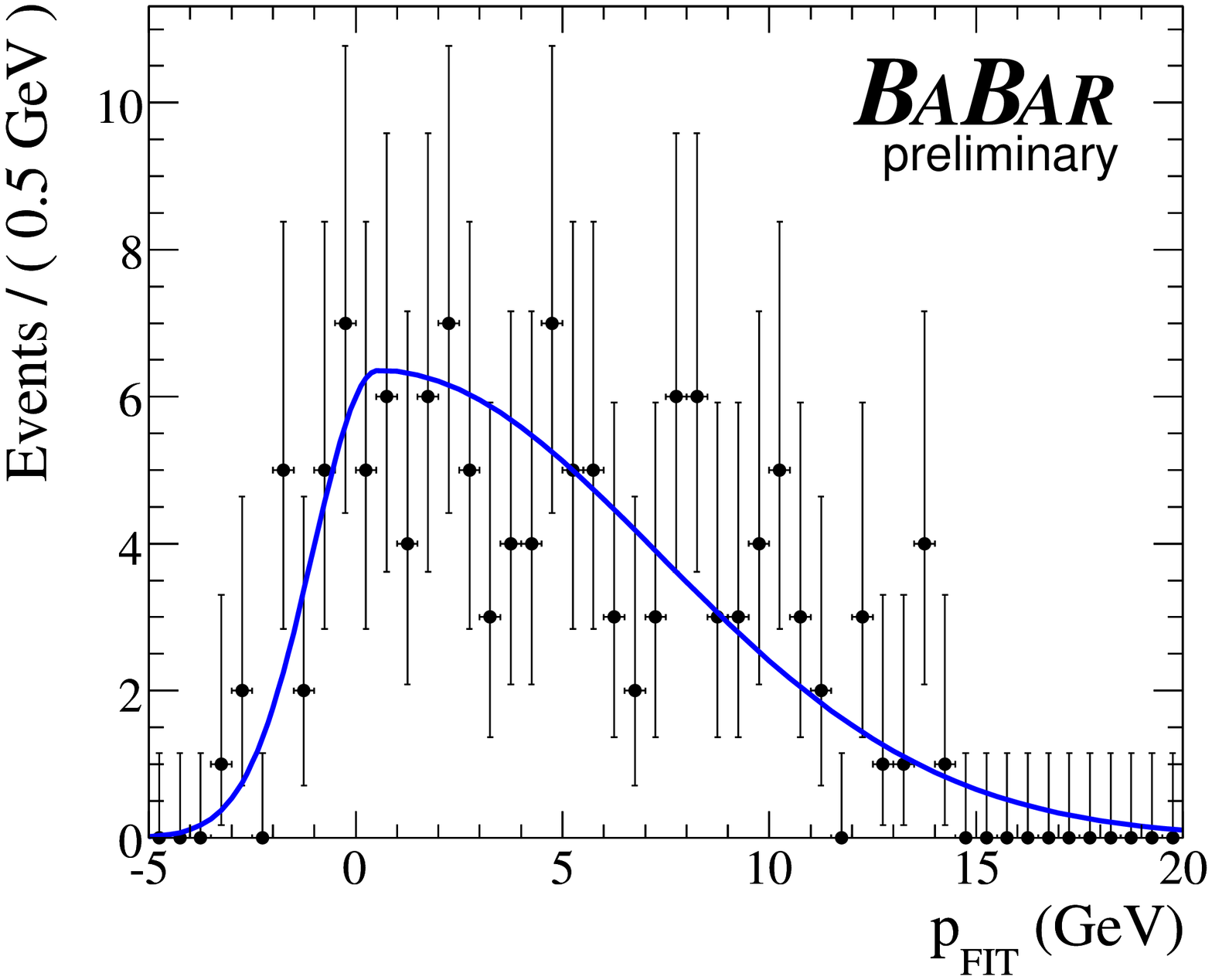}
\caption{ Distribution of $p_{\rm{FIT}}$  with the fit superimposed: signal (top left), $b \bar{b}$ (top right) and $c \bar{c}$ + $uds$ + $\tau^+ \tau^-$ (bottom). }
\label{fig:pfit_fit} 
\end{center}
\end{figure}

The two-body kinematics of this decay is now exploited by combining $p^*$ and $p_{\rm{CM}}$
in a second Fisher discriminant in order to discriminate against the remaining
semileptonic $b \bar{b}$ background events.
Signal and background yields are obtained from a Maximuum Likelihood Fit using the 
Fisher output $p_{\rm{FIT}}$.
We parameterize signal MC with the sum of two Gaussians.
As $b \bar{b}$ events and continuum $q \bar{q}$ and $\tau^+ \tau^-$ events 
are two background samples with different $p_{\rm{FIT}}$ Probability Density Functions (PDFs), 
we parameterize them separately and construct a summed background PDF 
with relative normalizations fixed from simulated events. Both of them are parameterized with a Gaussian
function with a different sigma for value above and below the peak (bifurcated
Gaussians).
Table~\ref{tab:fit_par} shows the fixed parameterization of signal and backgrounds PDFs,
with purely statistical uncertainties arising from the size of the simulated 
datasets used to obtain the parameterizations. The $p_{\rm{FIT}}$ distributions for simulated
signal and background events are shown in Figure~\ref{fig:pfit_fit}. Only the signal yield and 
the yield of the sum of all backgrounds are free parameters in the fit.

\begin{table}[htb!]
\caption{ $p_{\rm{FIT}}$ distribution parameterization for signal MC (left), $b \bar{b}$ (center) and $uds$ + $c \bar{c}$ +   $\tau^+ \tau^-$ (right).
The two summed Gaussians have parameters ($\mu_{\rm{core}}$,$\sigma_{\rm core}$) and ($\mu_{\rm{tail}}$,$\sigma_{\rm tail}$) respectively and 
$f_{\rm{core}}$ is the relative fraction of the core Gaussian.
The bifurcated Gaussian has $\mu_{\rm{core}}$ as mean and $\sigma_L$ and $\sigma_R$ as left and right $\sigma$ respectively. }
\begin{center}
\begin{tabular}{cccc}
\hline \hline
Parameter                      & Signal MC     &  $b \bar{b}$   & $uds$ + $c \bar{c}$ +   $\tau^+ \tau^-$  \\
\hline \hline
$\mu_{\rm core}$               & 5.32 $\pm$ 0.03 & -2.2 $\pm$ 0.3 & 0.5 $\pm$ 1.1 \\
$\sigma_{\rm core (L)}$        & 2.43 $\pm$ 0.04 & 0.7 $\pm$ 0.2  & 1.6 $\pm$ 0.7 \\
$\mu_{\rm{tail}}$        & 4.91 $\pm$ 0.03 &  - & -\\    
$\sigma_{\rm{tail} (R)}$ & 1.39 $\pm$ 0.04 & 3.6 $\pm$ 0.3 & 6.8 $\pm$ 0.9 \\
$f_{\rm core}$                 & 0.54 $\pm$ 0.03 &  - & -\\ 
\hline
\end{tabular}
\end{center}
\label{tab:fit_par}
\end{table}

\section{SYSTEMATIC STUDIES}
\label{sec:Systematics}

\begin{figure}[htb!]
\begin{center}
\includegraphics*[width=0.40\textwidth]{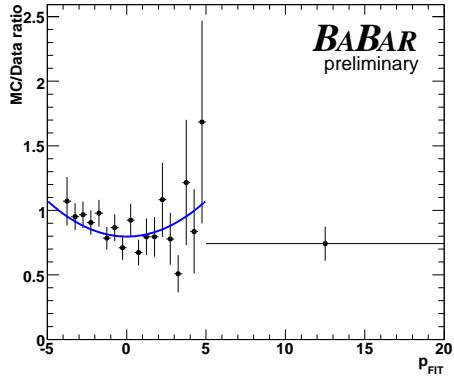}
\caption{MC/Data ratio of $p_{FIT}$ distribution on $\Delta E$ and $m_{ES}$ sideband ($\Delta E >$ 0 GeV and 5.2 GeV/c$^2$ $ < m_{ES} < $5.246 GeV/c$^2$). }
\label{fig:ratio} 
\end{center}
\end{figure}

To set an upper limit on the \bmunu branching fraction we evaluate systematic 
uncertainties in the number of \Bpm in the sample, the signal efficiency  and the signal 
yield. 
\begin{itemize}

\item The number of \Bpm mesons in the on-peak data sample is estimated to be 447 $\times$ 10$^6$ with an uncertainty of 1.1\%
 estimated studying $\mu \mu$ pairs events~\cite{Aubert:2002hc}. 

\item The uncertainty in the signal efficiency includes the muon candidate selection (particle identification, 
tracking efficiency and Fisher requirement) as well as the reconstruction efficiency of the tag \B. 
The muon identification efficiency systematic is evaluated using control samples derived
from the \babar\ data, which are weighted to reproduce the kinematic distribution 
of the muon signal candidate.
Comparing the cumulative signal efficiency obtained with and without this weight, 
a total discrepancy of 2.9$\%$ is found and this value is taken as the muon ID systematic uncertainty.

Charge conservation is imposed on  $\tau$ decays, which must proceed via an odd number of tracks and thus
the number of events with a missing track can be used to evaluate the uncertainty associated
with the tracking efficiency and the relative correction factor.
The systematic uncertainty per track and the correction factor are taken in quadrature to give the total
tracking efficiency uncertainty of 0.4$\%$ per track.

In order to evaluate the systematic uncertainty associated with the requirements on the Fisher discriminants,
we take the ratio between data and simulated events Fisher discriminant distributions 
in the $\Delta E$ and $m_{ES}$ sidebands $\Delta E > 0$ GeV and 5.2 GeV/c$^2$ $ < m_{ES} <$ 5.246 GeV/c$^2$
for each different data-taking period. We fit the data/MC 
ratio for each run with a linear function. The mean weighted by the errors of slopes and intercepts
returns a linear function consistent with a data/MC unitary ratio in the full Fisher range.
We take the uncertainty of 1.5$\%$ on the averaged intercept as the systematic
error on the Fisher discriminant cut.

The tag \B reconstruction has been studied with a control sample of 
$B^+\rightarrow D^{(*)0}\pi^+$ events, where the $D$ is reconstructed into $\bar{D}^{0} \rightarrow K^+ \pi^-$ 
and $D^{0} \rightarrow K^-  \pi^+ $  and the $D^*$ into $D^{*0} \rightarrow D^0 \gamma$ or 
$D^{*0} \rightarrow D^0 \pi^0$.
This is also a two-body decay so it is topologically very similar to our 
signal. Once reconstructed, the pion can be treated as if it were the signal muon and the $D^{(*)0}$ decay products 
are ignored to simulate the neutrino. The tag \B is then reconstructed in the control sample as it would be for
signal. 
We compare the efficiencies for our tag \B selection cuts in the $B^+\rightarrow D^{(*)0}\pi^+$
data and MC to quantify any data/MC disagreements that may affect the signal efficiency. We find a data/MC
discrepancy on $B^+\rightarrow D^{(*)0}\pi^+$ control sample of 2.7$\%$ and assign this as the signal 
efficiency uncertainty arising from the tag \B selection.
A summary of the systematic uncertainties in the signal efficiency is given in Table~\ref{tab:systematics_eff}. 
The final signal efficiency is thus 4.64 $\pm$ 0.19 \%.    
\begin{table}[!htb]
 \caption{ Contributions to the systematic uncertainty on the signal efficiency. }
 \begin{center}
 \begin{tabular}{cc} 
\hline \hline 
   Source                         & Relative Error \\
\hline \hline
    muon identification                  & 2.9\%         \\
    tracking efficiency                  & 0.4\%         \\
    tag \B reconstruction          & 2.7\%         \\
    Fisher selection                    & 1.5\%           \\
\hline
   Total                                 & 4.2\%         \\
\hline 
 \end{tabular}
 \end{center}
 \label{tab:systematics_eff}
\end{table}

\item The fit parameters are extracted from MC and are kept fixed in the final fit to extract the yields. These
parameters are affected by
an uncertainty due to the MC statistics, which is considered as  a source of systematic
uncertainty.
In order to evaluate it, the final fit has been repeated 500 times for each background and signal PDF 
parameter. We randomly generate the PDF parameters  assuming Gaussian errors and 
taking into account all the correlations between them.
We perform a Gaussian fit to the distribution of the number of signal events for each parameter, 
take the fitted sigma  as the systematic
uncertainty and sum in quadrature. The total systematic uncertainty in the signal yield
from all signal and background PDFs parameterization is 13 events.

We take into account possible discrepancies in the shape of the $p_{\rm{FIT}}$ background distribution
in data and simulated events using again the simulated events over data ratio in the $\Delta E$ and $m_{ES}$
sidebands $\Delta E > 0$ GeV and 5.2$ < m_{ES} <$ 5.246 GeV/c$^2$.
Given the low statistics available 
for values of $p_{\rm{FIT}}$ above 5, we parameterize this ratio with a parabolic function in the region
(-5;5) and take the constant value of the ratio in the region (5;20). The ratio with the parabolic
fit superimposed is shown in Figure~\ref{fig:ratio}. We repeat the final fit to the data 500 times
for each parabola parameter and the value of the constant. We generate each parameter according to a
Gaussian distribution centered at its mean value
and having a sigma equal to its error, taking into account all correlations between different
parameters. We weight the dataset by the generated ratio and repeat the fit. A Gaussian
fit to the distribution of number of fitted signal events for each parameter is performed and the
sigma of the the Gaussian fit is taken as systematic uncertainty and summed in quarature. The total
systematic uncertainty from this procedure is 7 events.
A summary of all systematic uncertainties in the fitted signal yield is provided in Table~\ref{tab:systematics_yield}.
\end{itemize}
The total systematic uncertainty is 1.1$\%$ on the number of $B^+ B^-$ pairs, 4.2 $\%$ on signal efficiency and 15 events on the signal yield. 

\begin{table}[!htb]
 \caption{ Contributions to the systematic uncertainty on the signal yield. }
 \begin{center}
 \begin{tabular}{cc} 
\hline \hline 
   Source                       & Total Error\\
\hline \hline
     PDF parameters               & 13   \\
     Data/MC agreement            &  7    \\
\hline
   Total                        & 15    \\
\hline 
 \end{tabular}
 \end{center}
 \label{tab:systematics_yield}
\end{table}

\section{RESULTS}
\label{sec:Results}

\begin{figure}[htb!]
\begin{center}
\includegraphics*[width=0.50\textwidth]{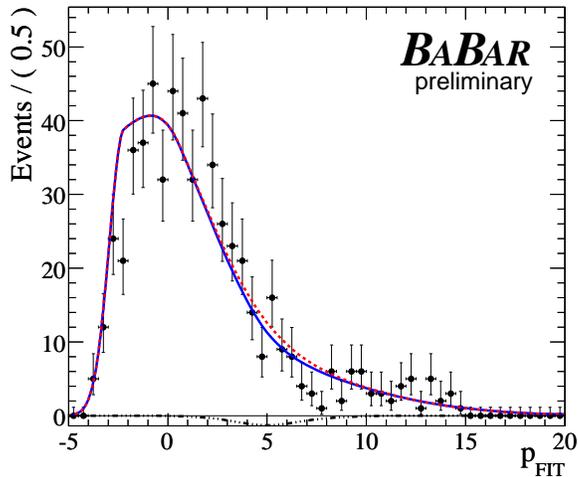}
\caption{Final fit to the data $p_{\rm{FIT}}$ distribution: full blue line is the total distribution, dashed red line
is background distribution, dashed-dotted black line is signal distribution.}
 \label{fig:unblind} 
\end{center}
\end{figure}

  From the data we extract  -12 $\pm$  15 signal events and  600 $\pm$ 29 background events. 
  We expect 10 events from MC assuming
  a SM branching fraction $\BR(B^{\pm} \rightarrow \mu^{\pm} \nu_{\mu} )= 4.7 \times 10^{-7}$. 
  The signal yield extracted corresponds to a central value  
$\BR(B^{\pm}\rightarrow\mu^{\pm}\nu_{\mu}) = ( -5.7 \pm 7.1 (\rm stat) \pm 6.8 (\rm syst) ) \times 10^{-7} $.
Figure~\ref{fig:unblind} shows the data points with the final fit superimposed. 

Given the number of fitted signal events, the signal efficiency and including all systematic
uncertainties, we find the Bayesian UL assuming a flat prior for the branching fraction up to a maximum of $\BR(B^{\pm} \rightarrow \mu^{\pm} \nu_{\mu} )= 3 \times 10^{-6}$
to be
\begin{eqnarray}
 \BR(\bmunu) < 1.3 \times 10^{-6}\nonumber
\end{eqnarray}
at the 90\% confidence level. The 95\% Bayesian UL is  $\BR(\bmunu) < 1.6 \times 10^{-6}$.
These results are more restrictive than previous measurements
from \babar\ ~\cite{Aubert:2004yu} and Belle~\cite{Satoyama:2006xn}. 

\section{ACKNOWLEDGMENTS}
\label{sec:Acknowledgments}

We are grateful for the 
extraordinary contributions of our \pep2\ colleagues in
achieving the excellent luminosity and machine conditions
that have made this work possible.
The success of this project also relies critically on the 
expertise and dedication of the computing organizations that 
support \babar.
The collaborating institutions wish to thank 
SLAC for its support and the kind hospitality extended to them. 
This work is supported by the
US Department of Energy
and National Science Foundation, the
Natural Sciences and Engineering Research Council (Canada),
the Commissariat \`a l'Energie Atomique and
Institut National de Physique Nucl\'eaire et de Physique des Particules
(France), the
Bundesministerium f\"ur Bildung und Forschung and
Deutsche Forschungsgemeinschaft
(Germany), the
Istituto Nazionale di Fisica Nucleare (Italy),
the Foundation for Fundamental Research on Matter (The Netherlands),
the Research Council of Norway, the
Ministry of Education and Science of the Russian Federation, 
Ministerio de Educaci\'on y Ciencia (Spain), and the
Science and Technology Facilities Council (United Kingdom).
Individuals have received support from 
the Marie-Curie IEF program (European Union) and
the A. P. Sloan Foundation.


\end{document}